\documentclass[letterpaper, 10 pt, conference]{cssconf}  % Comment this line out
\IEEEoverridecommandlockouts                              % This command is only
\usepackage{latexsym}
\usepackage{amsmath}
\usepackage{amssymb}
\usepackage{amsfonts}
\usepackage{graphicx}
\usepackage{epsfig}
\usepackage{array}

\newtheorem{theorem}{Theorem}
\newtheorem{lemma}{Lemma}
\newtheorem{ExampleDef}{Example}[section]
\newcommand{\Example}[3]{
  \begin{list}{}{
      \setlength{\leftmargin}{0em}}     % Indent everything by this amount
    \item                               % Group everything in one item
    \small                              % Use a smaller font size
    \begin{ExampleDef} \rm              % Theorems are italic - select roman
      {\bf \hspace{-1ex} #1}           % The name, use \\[1ex] to break line
      #2                                % The actual stuff
      \hfill {\large \boldmath $\Box$}  % The box
      \label{ex:#3}                      % Label the example
    \end{ExampleDef}
  \end{list}}
\title{\LARGE \bf
Adaptive Synchronization in Coupled Dynamic Networks}

\author{Wei Wang and Jean-Jacques E. Slotine % <-this % stops a space
\thanks{W. Wang is with the Department of Mechanical Engineering,
        Massachusetts Institute of Technology, 77 Mass. Av., Cambridge, MA 02139 USA
        {\tt\small wangwei@mit.edu}}%
\thanks{J.J.E. Slotine is with the  Department of Mechanical Engineering
and the Department of Brain and Cognitive Sciences,
        Massachusetts Institute of Technology, 77 Mass. Av., Cambridge, MA 02139 USA
        {\tt\small jjs@mit.edu}}%
}

\begin{document}

\maketitle
\thispagestyle{empty}
\pagestyle{empty}

%%%%%%%%%%%%%%%%%%%%%%%%%%%%%%%%%%%%%%%%%%%%%%%%%%%%%%%%%%%%%%%%%%%%%%%%%%%%%%%%
\begin{abstract}
This paper studies synchronization in coupled nonlinear dynamic
networks with unknown parameters.  Adaptation can be added to one or
several elements in the network, while preserving the global
synchronization conditions derived in~\cite{wei03-2, wei03-1}. This
implies that new nodes can be added to the network without prior
knowledge of the individual dynamics, and that nodes in an existing
network have the ability to recover dynamic information if temporarily
lost. In addition, when the individual elements feature sufficiently
rich stable dynamics , as e.g. in the case of oscillators, then
adaptation actually leads to an exact estimation of the unknown
parameters.  Different kinds of ``leaders'' are also discussed in this
context - one type of leader can specify overall trajectories for the
network, while another can concurrently specify dynamic parameters.
\end{abstract}

%%%%%%%%%%%%%%%%%%%%%%%%%%%%%%%%%%%%%%%%%%%%%%%%%%%%%%%%%%%%%%%%%%%%%%%%%%%%%%%%
\section{INTRODUCTION}

While the study of synchronization of coupled dynamic systems has a
long history~\cite{pikovsky, strogatz03}, it is still an extremely
active research topic~\cite{d'andrea, kumar, jadbabaie03, jadbabaie04,
leonard01, olfati_1, pecora90, vicsek02, malsburg81}. In two recent
papers~\cite{wei03-2, wei03-1}, we proposed a new theoretical analysis
tool, which we called {\em Partial Contraction Theory}, which is based
on {\em Contraction} concept~\cite{winni98}. We used Partial
Contraction Theory to analyze the collective behaviors of dynamic
networks with arbitrary size and general connectivity.
Synchronization conditions were derived for networks with
diffusion-like couplings, which essentially require the network to be
connected, the maximum eigenvalue of the uncoupled Jacobian matrix to
be upper bounded, and the coupling strengths to exceed a threshold.

In this paper, we extend the results in~\cite{wei03-2, wei03-1} to
coupled networks with adaptation.  We show that synchronization still
occurs under very similar conditions if adaptation is added to one or
more than one elements in a network. Estimated parameters will
converge to real values if the stable system behaviors are
sufficiently rich or persistently exciting, which is the case when the
individual elements are oscillators, for instance.  This result is
conceptually new as it differs with most existing adaptation models,
in the sense that the adaptation here only adds to a small part of the
network, and the target behavior is synchronization rather than an
explicit desired trajectory. Imagine for instance that some elements
in a network lose the normal values of some parameters. They can
recover the losing information from the rest of the network, as long
as there are still elements holding this information. The nodes
holding the real parameters can be considered ``knowledge-based
leaders''. They differ with the usual ``power-based'' leaders or
virtual leaders such as those in e.g. ~\cite{jadbabaie03,leonard01}
and~\cite{wei03-2, wei03-1}, which specify desired trajectories for
the network by unidirectionally coupling to it. With adaptation, we
can also add new nodes into a network without knowing the dynamic
properties of the nodes already in the network.
 
Section~\ref{sec:review} briefly reviews Contraction and Partial
Contraction Theories, and the main results we derived for
synchronization.  Adaptation for two coupled systems is studied in
Section~\ref{sec:adaptation-two}, and the results are extended to
general coupled networks in Section~\ref{sec:adaptation-network}. Brief
concluding remarks are offered in Section~\ref{sec:conclusion}.

%%%%%%%%%%%%%%%%%%%%%%%%%%%%%%%%%%%%%%%%%%%%%%%%%%%%%%%%%%%%%%%%%%%%%%%%%%%%%%%%
%
\section{Synchronization in Coupled Networks} \label{sec:review}

\subsection{Graph Theory Preliminaries} \label{sec:graph_theory}
Let us first introduce some basic Graph Theory concepts~\cite{fiedler,
godsil, mohar91} which will be used in the rest of the paper.

A graph $G$ is composed of a set of $n$ nodes and a set of $\tau $
links.  If there is a direction of flow associated with each link, $G$
is called a directed graph, otherwise it is undirected.  $G$ is
connected if any two nodes inside are linked by a path.  The
components of the {\it adjacency matrix} ${\bf A}(G) \in \mathbb{R}^{n
\times n}$ are defined as $[{\bf A}]_{ij} = 1$, for an undirected
graph if there is a link connecting nodes $j$ and $i$, and for a
directed graph if there is a link from node $j$ to node $i$, and as
$[{\bf A}]_{ij}=0$ otherwise. The {\em valency matrix} ${\bf D}(G) \in
\mathbb{R}^{n \times n}$ is a diagonal matrix with $[{\bf D}]_{ii} =
\sum_{j=1}^n a_{ij}$.  The matrix ${\bf L}(G) = {\bf D} - {\bf A}$ is
the {\em Laplacian matrix}, which is symmetric and positive
semi-definite if $G$ is undirected. Its second minimum eigenvalue
$\lambda_2({\bf L})\ $ is the {\em algebraic connectivity}, which is
zero if and only if $G$ is not connected. The first eigenvalue is
always zero, corresponding to the eigenvector $[1,1,\ldots,1]^T$.

Assign an arbitrary orientation $\sigma$ to an undirected graph $G$. 
We get the {\em incidence matrix} 
$= {\bf D}(G^{\sigma}) \in \mathbb{R}^{n \times \tau}$. 
For each oriented link $k$ which starts from node $i$ and ends at node $j$, 
$[{\bf D}]_{ik} = 1$ and $[{\bf D}]_{jk} = -1$. All the other entries
of ${\bf D}$ are equal to $0$. Moreover,
$$
{\bf L}(G) = {\bf D}(G^\sigma)\ {\bf D}^T(G^\sigma)
$$

If the graph is weighted, we have the weighted Laplacian matrix
$$
{\bf L}_{\bf \mathcal{K}}\ =\ {\bf D}\ {\bf \mathcal{K}} \ {\bf D}^T
$$
where $\ {\bf \mathcal{K}} \in \mathbb{R}^{\tau \times \tau}\ $ is
a diagonal matrix with the $k^{\rm{th}}$ diagonal entry 
$\ [{\bf \mathcal{K}}]_k = {\bf K}_{ij}\ $ corresponding to the weight of the 
$k^{\rm{th}}$ link. If ${\bf K}_{ij} \in \mathbb{R}^{m \times m}$
is a matrix, ${\bf \mathcal{K}}$ is block diagonal. Similarly
${\bf D}$ has block entries ${\bf I}$, $-{\bf I}$ and ${\bf 0}$.  

\subsection{Contraction and Partial Contraction Theories} \label{sec:theories}
In two recent papers~\cite{wei03-2, wei03-1}, we studied
synchronization behaviors of coupled dynamic networks. Here we briefly
introduce the theoretical analysis tools.

{\bf Contraction Theory}~\cite{winni98}:
Consider a nonlinear system
$
\dot{{\bf x}} = {\bf f}({\bf x},t)
$
where we assume ${\bf f}({\bf x},t)$ is continuously differentiable. 
Consider a virtual displacement $\delta {\bf x}$ between two neighboring solution 
trajectories. We have
$$
 \frac{d}{dt} ({\delta {\bf x}}^T {\delta {\bf x}})\ =\ 
  2 \ {\delta {\bf x}}^T \frac{\partial {\bf f}}{\partial {\bf x}} \ {\delta {\bf x}}
 \ \le\ 2\ \lambda_{max}\ {\delta {\bf x}}^T {\delta {\bf x}}
$$
where $\ \lambda_{max}({\bf x},t)\ $ is the largest eigenvalue of the symmetric part of 
the Jacobian matrix $\ \partial {\bf f} / \partial {\bf x}\ $. Hence, if
$\ \lambda_{max}({\bf x},t)\ $ is uniformly strictly negative, any
infinitesimal length $\ \| \delta {\bf x} \|\ $ converges exponentially to
zero. By path integration at fixed time, this implies in turn that all
the solutions converge exponentially to a single trajectory, independently 
of the initial conditions.

More generally, consider a coordinate transformation
$
{\delta {\bf z}} = {\bf \Theta} \delta {\bf x}
$
where ${\bf \Theta} ({\bf x},t)$ is a uniformly invertible square matrix.
We have
$$
\frac{d}{dt} ({\delta {\bf z}}^T {\delta {\bf z}})\ =\
2 \ {\delta {\bf z}}^T \ ( \dot{{\bf \Theta}} + {\bf \Theta} 
\frac{\partial {\bf f}}{\partial {\bf x}} ) {\bf \Theta}^{-1} \ \delta {\bf z}
$$ 
so that exponential convergence of $\| \delta {\bf z} \|$ to zero is guaranteed
if the  {\em generalized Jacobian matrix}
$$
{\bf F}=(\dot{\bf \Theta} + {\bf \Theta} 
\frac{\partial {\bf f}}{\partial {\bf x}} ) {\bf \Theta}^{-1}
$$ 
is uniformly negative definite. Again, this implies in turn that all
the solutions of the original system
converge exponentially to a single trajectory, independently of the
initial conditions. Such a system is called 
{\em contracting}~.

{\bf Partial Contraction Theory} \cite{wei03-2,wei03-1}:
Consider a nonlinear system in the form
$
\dot{\bf x}\ = \ {\bf f}({\bf x},{\bf x},t)
$. Consider the virtual, observer-like auxiliary system
$$
\dot{\bf y}\ = \ {\bf f}({\bf y},{\bf x},t)
$$ 
and assume that this system is contracting. By construction, $\ {\bf y}(t) =
{\bf x}(t)$ is a particular solution of the ${\bf y}$-system. Thus,
if another particular solution of the ${\bf y}$-system verifies a
smooth specific property, then all trajectories of the ${\bf
x}$-system verify this property exponentially. Such a ${\bf x}$-system
is called {\it partially contracting}.

\subsection{Synchronization of Coupled Networks} \label{sec:syn-network}
In this section, we list two main results in~\cite{wei03-2, wei03-1}, both of 
which will be extended through the rest of the paper.

{\bf Two Coupled Systems}:
Consider the coupled systems
\begin{equation} \label{eq:two-systems}
\begin{cases}
 \ \dot{\bf x}_1\ =\ {\bf h}({\bf x}_1,t)\ +\ {\bf g}({\bf x}_1,{\bf x}_2,t) \\
 \ \dot{\bf x}_2\ =\ {\bf h}({\bf x}_2,t)\ +\ {\bf g}({\bf x}_1,{\bf x}_2,t)
\end{cases}
\end{equation}
where ${\bf x}_1$, ${\bf x}_2 \in \mathbb{R}^m$ are the state vectors, 
and ${\bf g}$ could be any combination of ${\bf x}_1, {\bf x}_2$ and $t$. 
\begin{theorem}\label{th:twoway} 
If function ${\bf h}$ in~(\ref{eq:two-systems}) is contracting based on a 
constant ${\bf \Theta}$, ${\bf x}_1$ and ${\bf x}_2$ will converge 
to each other exponentially, regardless of the initial conditions.
\end{theorem}

The proof is based on the auxiliary system
$$
\dot{\bf y}\ =\ {\bf h}({\bf y},t)\ +\ {\bf g}({\bf x}_1(t),{\bf x}_2(t),t) 
$$
which is contracting and has two particular solutions ${\bf y}={\bf x}_1$
and ${\bf y}={\bf x}_2$.

{\bf Coupled Networks}:
Consider a coupled network containing $n$ elements
\begin{equation} \label{eq:general-network} 
\dot{\bf x}_i =  {\bf f}({\bf x}_i,t) + 
\sum_{j \in {\mathcal N}_i} {\bf K}_{ji}\ ({\bf x}_j - {\bf x}_i)  
\ \ \ \ i=1,\ldots,n
\end{equation}
where ${\mathcal N}_i$ denotes the set of indices of the active links
of element $i$, and the couplings are symmetric positive definite, i.e.,
${\bf K}_{ji} =  {\bf K}_{ij}>0$..
\begin{theorem} \label{th:general_network}
All the elements within the coupled network~(\ref{eq:general-network}) 
will reach group agreement exponentially if
\begin{itemize}
\item the network is connected
\item $\lambda_{max}({\bf J}_{is})$ is upper bounded
\item the coupling strengths are strong enough
\end{itemize}
where ${\bf J}_{is}$ is the symmetric part of the Jacobian
$\frac{\partial {\bf f}({\bf x}_i,t)} {\partial {\bf x}_i}$.
\end{theorem}
The proof is based on the auxiliary system
\begin{eqnarray} \label{eq:auxiliary}
\dot{\bf y}_i =  {\bf f}({\bf y}_i,t) + 
\sum_{j \in {\mathcal N}_i} {\bf K}_{ji}\ ({\bf y}_j - {\bf y}_i)
     - \ {\bf K}_0\ \sum_{j=1}^{n} ({\bf y}_j{- \bf x}_j(t))  \nonumber
\end{eqnarray}
which is contracting for appropriate choices of the constant matrix
${\bf K}_0 > 0$. All solutions of~(\ref{eq:general-network}) thus
verify the smooth specific property $\ {\bf x}_1 = \cdots = {\bf x}_n\
$ exponentially.

In fact, we can express the synchronization condition more specifically as
\begin{equation} \label{eq:syn_condition}
\lambda_{m+1} ( {\bf L}_{\bf \mathcal{K}} )\ >\ \max_i \lambda_{max}({\bf J}_{is})
 \ \ \ \ \ \mathrm{uniformly}
\end{equation}
where ${\bf L}_{\bf \mathcal{K}}$ 
is the weighted Laplacian matrix with each weight corresponding to the coupling
strength of that link.

In addition, the coupling forces in a network can be more general,
such as uni-directional, positive semi-definite, or nonlinear.

Note that the condition~(\ref{eq:syn_condition}) is always true if 
${\bf J}_{is} \le 0$ and the network is connected. For example, assuming
${\bf f}({\bf x}_i,t)=0$, system~(\ref{eq:general-network}) represents
a schooling or flocking model~\cite{wei03-2} which provides an effective 
distributed solution for group agreement problem~\cite{jadbabaie03, leonard01, olfati_1} or 
rendezvous problem~\cite{ando, morse}.

%%%%%%%%%%%%%%%%%%%%%%%%%%%%%%%%%%%%%%%%%%%%%%%%%%%%%%%%%%%%%%%%%%%%%%%%%%%%%%%%
%
\section{Two Coupled  Systems with Adaptation} 
\label{sec:adaptation-two}
\begin{figure}[h]
\begin{center}
\epsfig{figure=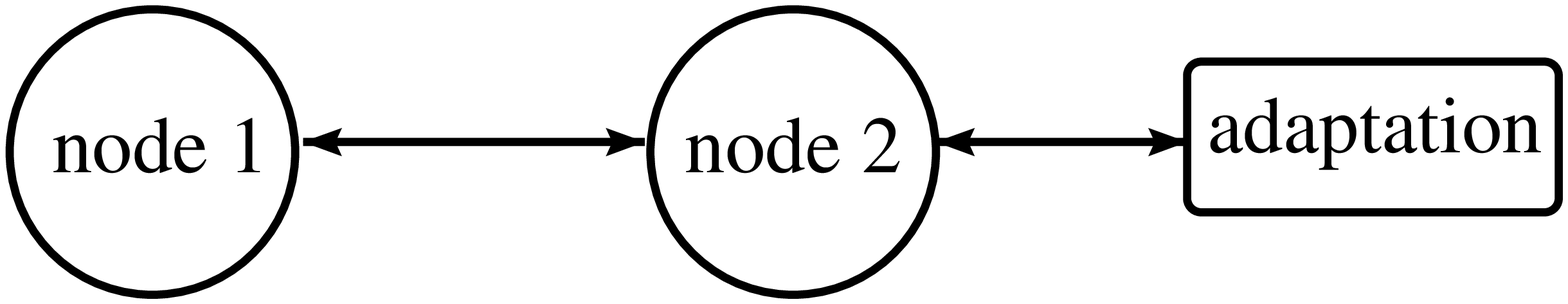,height=10mm,width=55mm}
\end{center}
\caption{ The structure of two coupled systems with adaptation.} 
\label{fig:adp-two}
\end{figure}

Consider two coupled systems as in~(\ref{eq:two-systems}), but assume
that a parameter vector ${\bf a}$ is unknown to the second system. To
guarantee state convergence, we generate an estimated parameter
$\hat{\bf a}$ through an adaptation mechanism. Specifically, the ${\bf
x}_2$ dynamics is replaced by
\begin{eqnarray} \label{eq:two-adaptation}
 \dot{\bf x}_2 &=& {\bf h}({\bf x}_2,\hat{\bf a},t)\ +\ {\bf g}({\bf x}_1,{\bf x}_2,t) \\
 \dot{\hat{\bf a}} &=& {\bf P} {\bf W}^T({\bf x}_2,t)\tilde{\bf x}, \ \ \ \
            \tilde{\bf x} = {\bf x}_1-{\bf x}_2 \nonumber 
\end{eqnarray}
with constant symmetric ${\bf P}>0$ and ${\bf W}({\bf x}_2,t)$ defined as
$$
{\bf h}({\bf x}_2,\hat{\bf a},t)\ =\ {\bf h}({\bf x}_2,{\bf a},t)\ +\ 
{\bf W}({\bf x}_2,t)\tilde{\bf a}
$$
with $\ \tilde{\bf a} = \hat{\bf a}-{\bf a}\ $. 
A similar adaptive technique was used in~\cite{winni98, winnithesis}, but
is generalized here in the sense that the couplings are bidirectional. 
The system structure is illustrated in Figure~\ref{fig:adp-two}.
\begin{theorem}\label{th:adaptation-two}
In system~(\ref{eq:two-adaptation}), $\tilde{\bf x}$ converges to $0$ 
asymptotically if ${\bf x}_1$ is bounded and ${\bf h}$ is contracting.
\end{theorem}

{\bf Proof:} Define the Lyapunov function
\begin{eqnarray*}
V &=& \frac{1}{2}\ (\ \tilde{\bf x}^T \tilde{\bf x}\ +\ 
                      \tilde{\bf a}^T {\bf P}^{-1} \tilde{\bf a}\ )\ >\ 0 \\
\dot{V} &=& \tilde{\bf x}^T\ (\ {\bf h}({\bf x}_1,{\bf a},t)\ -\ 
            {\bf h}({\bf x}_2,{\bf a},t)\ ) \\
        &=& \tilde{\bf x}^T\ \int_0^1 \frac{\partial {\bf h}}{\partial {\bf x}}
            ({\bf x}_2 + \chi \tilde{\bf x})d \chi\ \tilde{\bf x}\ \le\ 0
\end{eqnarray*} 
The boundedness of ${\bf x}_1$ implies that of ${\bf x}_2$. Assuming all the 
functions are smoothly differentiable, the boundedness of $\ddot{V}$ can be 
concluded since all the states including ${\bf a}$ are bounded.
According to Barbalat's lemma~\cite{jjsbook}, $\dot{V}$ and therefore $\tilde{\bf x}$ 
tends to $0$ asymptotically.  \hfill {\large \boldmath $\Box$}

Note that $\dot{\hat{\bf a}}$ also tends to $0$. Furthermore, since 
$\ddot{\tilde{\bf x}}$ is also bounded, we have the asymptotic convergence 
of $\dot{\tilde{\bf x}}$ to zero, which leads to the convergence of 
${\bf W}({\bf x}_2,t)\tilde{\bf a}$ to zero. In particular~\cite{jjsbook}, if 
$$
\exists\ \alpha>0, T>0, \ \ \forall t \ge 0\ \ 
\int_t^{t+T} {\bf W}^T {\bf W} dr \ \ge\ \alpha {\bf I}
$$
then $\tilde{\bf a}$ converges to zero asymptotically, too.

The boundedness of ${\bf x}_1$ is trivial if ${\bf g}={\bf g}({\bf x}_1, t)$, which
is a classical observer structure with ${\bf x}_1$ dynamics independent. 
If ${\bf g}={\bf g}({\bf x}_1, {\bf x}_2, t)$, we have
$$
\dot{\bf x}_1 = {\bf h}({\bf x}_1,{\bf a},t)\ +\ {\bf g}({\bf x}_1,{\bf x}_1-\tilde{\bf x},t)
              = {\bf e}({\bf x}_1,\tilde{\bf x},t)
$$
where $\tilde{\bf x}$ is bounded. Thus the boundedness of ${\bf x}_1$ is determined by the
Input-to-State Stability~\cite{khalil} of ${\bf e}$.

\begin{figure}[h]
\begin{center}
\epsfig{figure=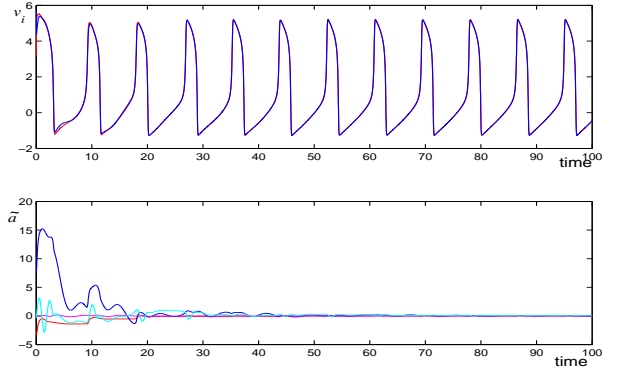,height=50mm,width=80mm}
\end{center}
\caption{Simulation result of Example~\ref{ex:adaptation-fn-two}. The
real parameters are $\ \alpha = 5.32, \beta = 3,
\gamma = 0.1\ $ and $\ I=20\ $. The coupling gain $\ k = 15\ $. The matrix
$\ {\bf P} = \mathrm{diag}\{ 0.6, 30, 0.002, 0.4 \}\ $. All initial
conditions are chosen arbitrarily. The plots are (a) states $v_i$ versus 
time, (b) estimator error $\tilde{\bf a}$ versus time. } 
\label{fig:adp-simulation}
\end{figure}

\Example{}{
Consider two coupled FitzHugh-Nagumo (FN) neurons~\cite{fitzhugh, murray, nagumo}, a
famous spiking neuron model,
\begin{eqnarray*} \label{eq:two-fn-adaptation}
& & \begin{cases}  
  \ \dot{v}_1 = v_1(\alpha - v_1)(v_1-1)-w_1+I+k(v_2-v_1)  \\  
  \ \dot{w}_1 = \beta v_1 - \gamma w_1 
\end{cases} \\
& & \begin{cases}  
  \ \dot{v}_2 = v_2(\hat{\alpha} - v_2)(v_2-1)-w_2+\hat{I}+k(v_1-v_2)  \\  
  \ \dot{w}_2 = \hat{\beta} v_2 - \hat{\gamma} w_2 
\end{cases}
\end{eqnarray*}
where $v$ is the membrane potential and $I$ external stimulation current. We have
${\bf x} = \left[ \begin{array}{cc} v & w \end{array} \right]^T$, 
${\bf a} = \left[ \begin{array}{cccc} \alpha & I & \gamma & \beta \end{array} \right]^T>0$ and
$$
{\bf h}({\bf x},{\bf a},t) = 
\left[ \begin{array}{c} v(\alpha - v)(v-1)-w+I-2kv \\ \beta v-\gamma w \end{array} \right]
$$
which is contracting if the coupling gain $k$ is larger than an 
explicit threshold~\cite{wei03-2}. The adaptive law is thus
$$
\dot{\hat{\bf a}}\ =\ {\bf P} {\bf W}^T \tilde{\bf x}
$$
with
$$
{\bf W}\ =\ \left[ \begin{array}{cccc} v_2^2-v_2 & 1 & 0 & 0 \\ 0 & 0 & -w_2 & v_2 
          \end{array} \right]
$$
Note that although the diffusion couplings are only based on variable $v$, full-state 
feedback is needed for adaptation in this case. 
See Appendix~\ref{ap:FN-neuron-proof} for the boundedness proof. The simulation result
is illustrated in Figure~\ref{fig:adp-simulation}.
}{adaptation-fn-two}

%%%%%%%%%%%%%%%%%%%%%%%%%%%%%%%%%%%%%%%%%%%%%%%%%%%%%%%%%%%%%%%%%%%%%%%%%%%%%%%%
%
\section{Adaptation in Coupled Networks} \label{sec:adaptation-network}
\subsection{Basic Results} \label{sec:basic-results}
Consider now the coupled network~(\ref{eq:general-network}).
We show that similar synchronization conditions as those in 
Theorem~\ref{th:general_network} can be derived if adaptation
is added. 
 
\begin{figure}[h]
\begin{center}
\epsfig{figure=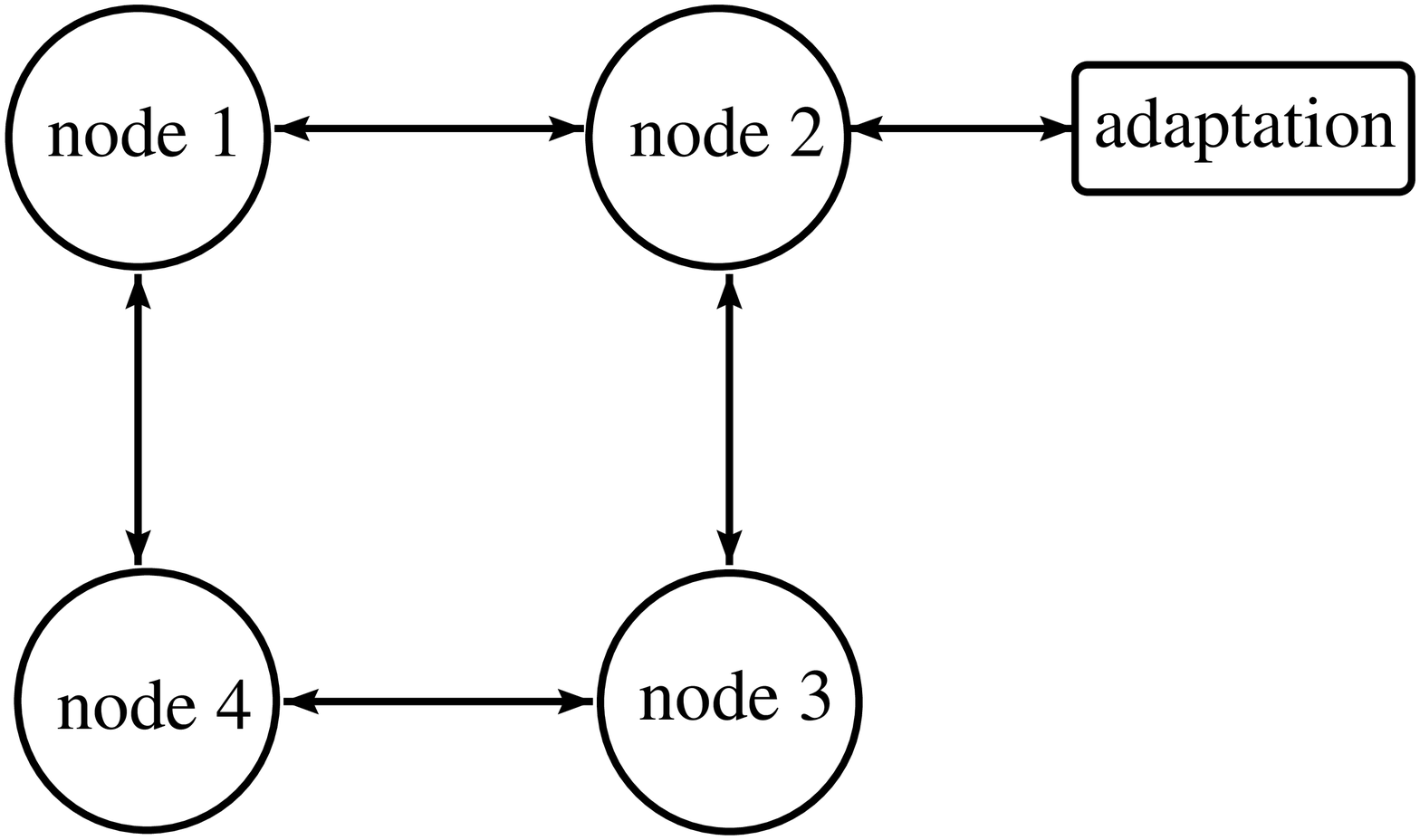,height=25mm,width=50mm}
\end{center}
\caption{Coupled network with adaptation}
\label{fig:adaptation-network}
\end{figure}

For simplicity, we assume that the couplings are bidirectional, which means
the corresponding graph is undirected, and 
$\ {\bf K}_{ji} = {\bf K}_{ij} = {\bf K}_k\ $ with $k$ denoting the $k^{th}$ link
$(i,j)$. We also assume that each coupling gain is symmetric
positive definite, that is, $\ {\bf K}_k^T= {\bf K}_k>0\ $.
Assume now that the uncoupled dynamics 
$\ {\bf f}({\bf x}_i, t) = {\bf f}({\bf x}_i, {\bf a}, t)\ $ 
contains a parameter set ${\bf a}$ which is unknown to an arbitrary 
node $\varsigma$. We then use the estimated parameter $\hat{\bf a}$ in 
$\varsigma$, with an adaptive law based on the local ``coupling forces''
\begin{equation} \label{eq:adaptive-law-network}
\dot{\hat{\bf a}}\ =\ {\bf P} {\bf W}^T({\bf x}_{\varsigma},t) 
 \sum_{j \in {\mathcal N}_{\varsigma}} {\bf K}_{j\varsigma}\ ({\bf x}_j - {\bf x}_{\varsigma})  
\end{equation}
where ${\bf P}$ and ${\bf W}({\bf x}_{\sigma},t)$ are the same as those defined 
in~(\ref{eq:two-adaptation}). The network structure is illustrated in 
Figure~\ref{fig:adaptation-network}. Note that the adaptation only uses
feedback from the connected neighbors.

To prove convergence, we define a Lyapunov function
$$
V = \frac{1}{2}\ (\ {\bf x}^T {\bf L}_{\bf \mathcal{K}} {\bf x}\ +\ 
                 \tilde{\bf a}^T {\bf P}^{-1} \tilde{\bf a}\ )
$$
where $\ {\bf L}_{\bf \mathcal{K}} = {\bf D} {\bf \mathcal{K}} {\bf D}^T \ $ 
is the weighted Laplacian matrix, and each weight corresponds to the coupling
strength of that link. ${\bf L}_{\bf \mathcal{K}}$ is symmetric positive 
semi-definite since ${\bf \mathcal{K}}$ is symmetric positive definite. 
We show that
\begin{eqnarray*}
\dot{V} 
&=& {\bf x}^T {\bf L}_{\bf \mathcal{K}} \dot{\bf x}\ +\ 
                 \tilde{\bf a}^T {\bf P}^{-1} \dot{\hat{\bf a}} \\
&=&   {\bf x}^T {\bf L}_{\bf \mathcal{K}}\ (\  
               \left[ \begin{array}{c} {\bf f}({\bf x}_1,{\bf a},t) \\ \ldots \\  
                {\bf f}({\bf x}_n,{\bf a},t) \end{array} \right]
        - {\bf L}_{\bf \mathcal{K}}{\bf x}\ ) \\
&=& {\bf x}^T\ {\bf D} {\bf \mathcal{K} \Lambda}{\bf D}^T\ {\bf x} - 
          {\bf x}^T\ {\bf L}_{\bf \mathcal{K}}^2\ {\bf x} \\
&=& {\bf x}^T\ (\ {\bf L}_{\bf \mathcal{K} \Lambda} - 
          {\bf L}_{\bf \mathcal{K}}^2\ )\ {\bf x}
\end{eqnarray*}
where
$$
{\bf L}_{\bf \mathcal{K} \Lambda}^T\ =\ {\bf L}_{\bf \mathcal{K} \Lambda}\ 
=\ {\bf D}\ ({\bf \mathcal{K}} {\bf \Lambda})_s\ {\bf D}^T
$$
The matrix ${\bf \Lambda} \in \mathbb{R}^{\tau \times \tau}\ $ is a block diagonal 
matrix with the $k^{\rm{th}}$ diagonal entry 
$$
[{\bf \Lambda}]_k = \int_0^1 \frac{\partial {\bf f}}{\partial {\bf x}}
            ({\bf x}_j + \chi ({\bf x}_i - {\bf x}_j))\ d \chi
$$
corresponding to the $k^{\rm{th}}$ link, which
has been assigned with an arbitrary orientation by incidence matrix ${\bf D}$, 
for instance from node $i$ to node $j$.
The matrix $({\bf \mathcal{K}} {\bf \Lambda})_s\ $, the symmetric part of 
$\ {\bf \mathcal{K}} {\bf \Lambda}\ $, is also a block diagonal matrix with
$k^{\rm{th}}$ diagonal entry $\ [({\bf \mathcal{K}} {\bf \Lambda})_s]_k = 
({\bf K}_k [{\bf \Lambda}]_k)_s\ $.
\begin{lemma} \label{lm:negative-semi-definite}
The matrix $\ {\bf L}_{\bf \mathcal{K} \Lambda} - {\bf L}_{\bf \mathcal{K}}^2\ $ is
negative semi-definite if
\begin{equation} \label{eq:negative-semi-definite}
\frac{\lambda_{m+1}^2 ( {\bf L}_{\bf \mathcal{K}} )}{\lambda_n({\bf L})} > 
\max_k \lambda_{max}({\bf K}_k [{\bf \Lambda}]_k)_s
\end{equation}
See Appendix~\ref{ap:negative-semi-definite-proof} for the proof.
\end{lemma}

\begin{lemma} \label{lm:eigenvectors}
Given any vector ${\bf x}=[{\bf x}_1^T, {\bf x}_2^T, \ldots, {\bf x}_n^T]^T$.
For a coupled network, if the condition~(\ref{eq:negative-semi-definite}) is true,
$$
{\bf x}^T\ (\ {\bf L}_{\bf \mathcal{K} \Lambda}-{\bf L}_{\bf \mathcal{K}}^2\ )\ {\bf x}
=0
$$ if and only if 
$\ {\bf x}_1 = {\bf x}_2 = \cdots = {\bf x}_n$.
See Appendix~\ref{ap:eigenvectors-proof} for the proof.
\end{lemma}

\begin{theorem}\label{th:adaptation-network}
For a connected dynamic network~(\ref{eq:general-network}) with 
adaptation~(\ref{eq:adaptive-law-network}) adding to an arbitrary node $\varsigma$,
the states of all the elements will converge together asymptotically if
the condition~(\ref{eq:negative-semi-definite}) is true and all the states are
bounded.
\end{theorem}

{\bf Proof:}
Similar to the proof of Theorem~\ref{th:adaptation-two}, if all the states are bounded, 
we can conclude the boundedness of $\ddot{V}$, which then leads to the asymptotically
convergence of $\dot{V}$ to zero if the condition~(\ref{eq:negative-semi-definite}) is 
true. With Lemma~\ref{lm:eigenvectors}, this implies immediately that
$\ {\bf x}_1 = {\bf x}_2 = \cdots = {\bf x}_n$. 
\hfill {\large \boldmath $\Box$}

{\it Remarks}:

$\bullet$\ Condition~(\ref{eq:negative-semi-definite}) in
Lemma~\ref{lm:negative-semi-definite} is in fact very similar to those
given in Theorem~\ref{th:general_network}. If $\lambda_{max}({\bf K}_k
[{\bf \Lambda}]_k)_s$ is positive, to guarantee synchronization one
needs a connected network, an upper bounded $\ \lambda_{max}({\bf K}_k
[{\bf \Lambda}]_k)_s$, and strong enough coupling strength ${\bf
\mathcal{K}}$.  This result thus implies that
adaptation~(\ref{eq:adaptive-law-network}) will not significantly
change the network's synchronization ability.

Assuming $m=1$ and all the coupling strengths 
$\ {\bf K}_k=\kappa\ $, a sufficient synchronization condition for a coupled network without 
adaptation is
$$
\kappa_1\ >\ \frac{\max_i {\bf J}_{is} }{\lambda_2({\bf L})}
$$
while it is
$$
\kappa_2\ >\ \frac{\lambda_n({\bf L})}{\lambda_2({\bf L})}\
\frac{\max_i {\bf J}_{is} }{\lambda_2({\bf L})}
$$
with adaptation. Here $\ {\bf J}_{is}= (\partial {\bf f} / \partial {\bf x}_i)_s$.
%A higher coupling-strength threshold $\kappa_2$ is caused partly by the choice of the function $V$.
%A lower value of $\kappa$ will still work in practice but it may not be able to guarantee
%non-oscillating convergence of $V$ to zero. To show this, let us consider a network without
%adaptation,
%\begin{eqnarray*}
%V &=& \frac{1}{2}\ {\bf x}^T {\bf L}_{\bf \mathcal{K}} {\bf x} 
% \ =\ \frac{1}{2}\ ({\bf x}-{\bf y})^T {\bf L}_{\bf \mathcal{K}} ({\bf x}-{\bf y}) \\
%  &\le& \frac{1}{2}\ \lambda_{max}( {\bf L}_{\bf \mathcal{K}} ) ({\bf x}-{\bf y})^T ({\bf x}-{\bf y})
%\end{eqnarray*}
%where $\ {\bf y}=[{\bf y}_1, {\bf y}_2, \ldots, {\bf y}_n]^T\ $ is a particular solution of the 
%auxiliary system and $\ {\bf y}_1 = {\bf y}_2 = \cdots = {\bf y}_n\ $
%since we assume all the subsystems ${\bf y}_i$ have the same initial conditions. The
%exponential convergence of $({\bf x}-{\bf y})^T ({\bf x}-{\bf y})$ to zero can be
%proved under the conditions in Theorem~\ref{th:general_network}. But it only guarantees
%convergence, not non-oscillating convergence, of $V$ to zero. 
%The difference is illustrated in Figure~\ref{fig:adp-network} through the comparison of 
%plots (b) and (d). 
%
%We can certainly choose other potential Lyapunov functions, such as
%$\ V= \frac{1}{2} {\bf x}^T {\bf L} {\bf x}$. The result will be slightly different.

$\bullet$\ When the coupling gains are only positive {\it
semi-definite}, extra restrictions have to be added to the uncoupled
system dynamics to guarantee globally stable synchronization,
similarly to the fixed-parameters result in~\cite{wei03-1}. See
Appendix~\ref{ap:eigenvectors-proof} for details.

$\bullet$\ Theorem~\ref{th:adaptation-network} requires the states to
be bounded. Boundedness can be shown following the same steps as we did in
Section~\ref{sec:adaptation-two}. In fact, since $\dot{V} \le 0$, we
know that $\tilde{\bf a}$ and $\forall k$, $\tilde{\bf x}_k = {\bf
x}_i - {\bf x}_j$ are bounded. Thus the boundedness of the states are
determined by the Input-to-State Stability of the system
$$
\dot{\bf x}_i =  {\bf f}({\bf x}_i,t) + {\bf u}_i
$$
Convergence of the estimated parameter set $\hat{\bf a}$ can also be concluded
with the same analysis as that in Section~\ref{sec:adaptation-two}.

$\bullet$\ The result of Theorem~\ref{th:adaptation-network} still
holds if the parameter set ${\bf a}$ is unknown to multiple nodes and
adaptations are added to these nodes simultaneously. This can be shown
using the Lyapunov-like function
$$
V = \frac{1}{2}\ (\ {\bf x}^T {\bf L}_{\bf \mathcal{K}} {\bf x}\ +\ 
                 \sum_i \tilde{\bf a}_i^T {\bf P}^{-1}_i \tilde{\bf a}_i\ )
$$ 
The nodes holding the real parameters can thus be considered as
``knowledge-based'' leaders, which are different from usual
``power-based'' leaders~\cite{jadbabaie03,leonard01,wei03-2, wei03-1},
which specify desired trajectories for the network by unidirectionally
coupling to it. At the limit all nodes could be adaptive, although
they may then converge to any odd parameter set $-$ while all states
will converge together, the desired individual behaviors (such as
oscillations) may not be preserved depending on initial
conditions. Note that both ``power'' and ``knowledge'' leaders may be
virtual.

$\bullet$\ Note that in Lemma~\ref{lm:eigenvectors} the ${\bf x}_i$
may actually tend to zero together. We should exclude this
possibility, and the possibility that any component of ${\bf
x}_i$ converges to zero, with dynamic analysis $-$ for instance by 
showing that zero is an unstable state.

\Example{}{Consider a group of FN neurons connected in a general network
\begin{equation*} 
\begin{cases}  
\displaystyle 
  \ \dot{v}_i = v_i(\alpha - v_i)(v_i-1)-w_i+I + \sum_{j \in {\mathcal N}_i} k_{ij}(v_j-v_i)  \\  
  \ \dot{w}_i = \beta v_i - \gamma w_i 
\ \ \ \ \ \ \ \ \ i=1,\ldots,n
\end{cases}
\end{equation*}
Assume that all the parameters are now unknown to the node $\varsigma$.
We add adaptation 
$$
\dot{\hat{\bf a}}\ =\ {\bf P} {\bf W}^T\ \sum_{j \in {\mathcal N}_\varsigma}\ 
\left[ \begin{array}{cc} k_{j\varsigma} & 0 \\ 
                              0 & k_{j\varsigma}  \end{array} \right]\
(\ \left[ \begin{array}{c} v_j \\  w_j  \end{array} \right] - 
   \left[ \begin{array}{c} v_\varsigma \\  w_\varsigma  \end{array} \right]\ )
$$
with ${\bf W}$ the same as that defined in Example~\ref{ex:adaptation-fn-two}.
Simulation results are illustrated in Figure~\ref{fig:adp-network}.
\begin{figure}[h]
\begin{center}
\epsfig{figure=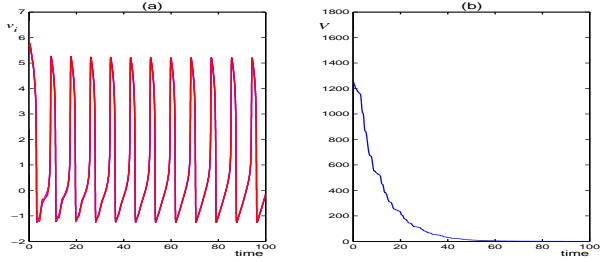,height=35mm,width=80mm}
\end{center}
\caption{ Simulation results of Example~\ref{ex:adaptation-fn-network}. The network
contains four FN neurons connected in a two-way ring as
Figure~\ref{fig:adaptation-network}. The real parameters and the coupling gains 
are the same as those in Figure~\ref{fig:adp-simulation}. The matrix
$\ {\bf P} = \mathrm{diag}\{ 0.06, 3, 0.0002, 0.04 \}\ $.  
All the initial conditions are chosen arbitrarily.
The plots are (a). states $v_i$ versus time, (b).$V$ versus time where,
$
V = \frac{1}{2}\ (\ {\bf x}^T {\bf L}_{\bf \mathcal{K}} {\bf x}\ +\ 
                 \tilde{\bf a}^T {\bf P}^{-1} \tilde{\bf a}\ )
$.} 
\label{fig:adp-network}
\end{figure}
}{adaptation-fn-network}

\subsection{Leader Combination} \label{sec:leader-combination}
We have mentioned that there can exist different leader roles in a
network, ones with power and ones with knowledge. And in fact, these
different types of leaders can co-exist. A leader guiding the
direction may use state measurements from its neighbors to adapt its
parameters to the values of the knowledge leaders.

Consider such a leader-followers network
\begin{eqnarray*}
\dot{\bf x}_0 &=&  {\bf f}({\bf x}_0,t)   \\
\dot{\bf x}_i &=&  {\bf f}({\bf x}_i,t) + \sum_{j \in {\mathcal N}_i} {\bf K}_{ji}({\bf x}_j - {\bf x}_i) 
                                        + \gamma_i {\bf K}_{0i} ({\bf x}_0 - {\bf x}_i) 
\end{eqnarray*}
where $i=1,\ldots,n$, ${\bf x}_0$ is the state of the group leader, $\gamma_i=0$ or
$1$, and ${\mathcal N}_i$ does not include the links with ${\bf
x}_0$. Adaptation can be added to any node(s) inside the network, such as
$$
\dot{\hat{\bf a}}_0\ =\ {\bf P}_0 {\bf W}^T({\bf x}_0,t) 
 \sum_{i=1}^n \gamma_i\ {\bf K}_{0i}\ ({\bf x}_i - {\bf x}_0)  
$$
for the leader, or
$$
\dot{\hat{\bf a}}_i\ =\ {\bf P}_i {\bf W}^T({\bf x}_i,t) 
 ( \sum_{j \in {\mathcal N}_i} {\bf K}_{ji}\ ({\bf x}_j - {\bf x}_i) +
  \gamma_i\ {\bf K}_{0i}\ ({\bf x}_0 - {\bf x}_i) \ ) 
$$
for the followers. To prove convergence, we define several Laplacian matrices

\noindent $\bullet$\ $\bar{\bf L}_{\bf \mathcal{K}}$, the weighted Laplacian of the 
followers network (excluding both the leader and the links from the leader).

\noindent $\bullet$\ $\vec{\bf L}_{\bf \mathcal{K}}$, the weighted Laplacian of the 
leader-followers network, which is not symmetric since we have uni-directional links.
Moreover,
$$
\vec{\bf L}_{\bf \mathcal{K}} = 
\left[ \begin{array}{cc} {\bf 0} & {\bf 0} \\                         
                   -{\bf b}  & {\bf C} \end{array} \right]
$$
where
$$
{\bf b} = \left[ \begin{array}{c} \vdots \\ \gamma_i\ {\bf K}_{0i} \\ \vdots  \end{array} \right] ,
\ \ \ \
{\bf C} = \bar{\bf L}_{\bf \mathcal{K}} + \mathrm{diag}\{ \gamma_i\ {\bf K}_{0i} \}
$$
Note that ${\bf C}$ is symmetric positive definite if the whole leader-followers network
is connected.

\noindent $\bullet$\ ${\bf L}_{\bf \mathcal{K}}$,  the weighted Laplacian of the 
leader-followers network if we consider it as an undirected graph. Thus,
$$
{\bf L}_{\bf \mathcal{K}} = {\vec{\bf L}_{\bf \mathcal{K}}}^T+
\left[ \begin{array}{cc} \sum_{i=1}^n \gamma_i {\bf K}_{0i} & {\bf 0} \\                         
                    -{\bf b}  & {\bf 0} \end{array} \right]
$$

Define
$$
V = \frac{1}{2}\ (\ {\bf x}^T {\bf L}_{\bf \mathcal{K}} {\bf x}\ +\ 
                 \sum_j \tilde{\bf a}_j^T {\bf P}^{-1}_j \tilde{\bf a}_j\ )
$$
where $j$ may be equal to any number from $0$ to $n$. We have
\begin{eqnarray*}
\dot{V} 
&=& {\bf x}^T {\bf L}_{\bf \mathcal{K}}\ (\  
               \left[ \begin{array}{c} {\bf f}({\bf x}_1,{\bf a},t) \\ \ldots \\  
                {\bf f}({\bf x}_n,{\bf a},t) \end{array} \right]
        - \vec{\bf L}_{\bf \mathcal{K}}{\bf x}\ ) \\
&=& {\bf x}^T\ (\ {\bf L}_{\bf \mathcal{K} \Lambda} - 
        {\vec{\bf L}_{\bf \mathcal{K}}}^T\ \vec{\bf L}_{\bf \mathcal{K}} \ )\ {\bf x}
\end{eqnarray*}
See Appendix~\ref{ap:leader-combination-proof} for the conditions for
${\bf L}_{\bf \mathcal{K} \Lambda} - {\vec{\bf L}_{\bf
\mathcal{K}}}^T\ \vec{\bf L}_{\bf \mathcal{K}}$ to be negative
semi-definite. Following the same proofs as in previous sections, this
then implies that all the states ${\bf x}_j$, $j=0,1,\ldots,n$
converge together.  Parameter convergence conditions are also the
same.

%%%%%%%%%%%%%%%%%%%%%%%%%%%%%%%%%%%%%%%%%%%%%%%%%%%%%%%%%%%%%%%%%%%%%%%%%%%%%%%%
%
\section{Concluding Remarks} \label{sec:conclusion}
Coupled networks with adaptation are studied in this paper. We showed
that synchronized behaviors can be achieved under similar conditions
as we derived in~\cite{wei03-2, wei03-1}, which also guarantees
parameter convergence if the stable system behaviors are sufficiently
rich. Different kinds of leaders may coexist in the network.  Current
work includes the applications of adaptive network, and the
investigation of coupled networks with distributed controllers.

\vspace{1.0em}

%\appendix
%%%%%%%%%%%%%%%%%%%%%%%%%%%%%%%%%%%%%%%%%%%%%%%%%%%%%%%%%%%%%%%%%%%%%%%%%%%%%%%%

\section{Appendices}

\subsection{Boundedness of Coupled FN Neurons} 
\label{ap:FN-neuron-proof}
For notation simplicity, define $u=I+k(v_2-v_1)$ and $\bar{w}=w/\sqrt{\beta}$. 
The dynamics of the first neuron changes to 
\begin{equation} \label{eq:input-to-state}
\begin{cases}  
  \ \dot{v}_1 = v_1(\alpha - v_1)(v_1-1) - \sqrt{\beta} \bar{w}_1 + u  \\  
  \ \dot{\bar{w}}_1 = \sqrt{\beta} v_1 - \gamma \bar{w}_1  
\end{cases}
\end{equation}
Define $\ U = \frac{1}{2}\ (\ v_1^2\ +\ \bar{w}_1^2\ ) $,
$$
\dot{U} = -(v_1-\alpha)(v_1-1)v_1^2\ -\ \gamma \bar{w}_1^2\ +\ u v_1
$$
Since $u$ is bounded, there must exist a large but bounded number $v_0>0$, 
$\forall\ |v_1|>v_0$, $\dot{U}<0$. We denote the region $|v_1| \le v_0$ as $\Omega$. 

If the system~(\ref{eq:input-to-state}) starts inside $\Omega$, since
the dynamics of $\bar{w}_1$ is linear and strictly stable, $v_1$ and
$\bar{w}_1$ are always bounded as long as the system stays inside
$\Omega$. In fact,
\begin{eqnarray*}
|\bar{w}_1(t)| &=& |(\ \bar{w}_1(0) + \int_0^t \sqrt{\beta} v_1(t) e^{\gamma t} dt\ )\ e^{-\gamma t}|  \\
            &\le & |\bar{w}_1(0)| e^{-\gamma t}\ +\ v_0 \frac{\sqrt{\beta}}{\gamma} (1-e^{-\gamma t})
\end{eqnarray*}
Thus, for any initial condition $|\bar{w}_1(0)| > v_0 \frac{\sqrt{\beta}}{\gamma}$, we have
$|\bar{w}_1(t)| \le |\bar{w}_1(0)|$, which implies that the bound of $|\bar{w}_1(t)|$ inside 
$\Omega$ is $\max (v_0 \frac{\sqrt{\beta}}{\gamma}, |\bar{w}_1(0)|)$.

\begin{figure}[h]
\begin{center}
\epsfig{figure=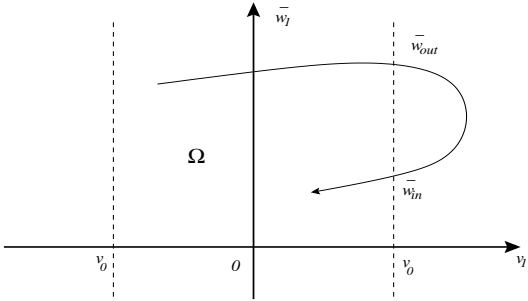,height=40mm,width=70mm}
\end{center}
\caption{Illustration of a solution trajectory of the system~(\ref{eq:input-to-state}) leaving
and re-entering the region $\Omega:\ |v_1| \le v_0$. } 
\label{fig:adp-illustration}
\end{figure}

Suppose that at some moment, the system leaves $\Omega$ through
point $(v_0\ \bar{w}_{out})$. Since $\dot{U}<0$ outside $\Omega$,
$v_1^2(t)$ and $\bar{w}_1^2(t)$ will be both bounded by $v_0^2+\bar{w}_{out}^2$ 
until the system trajectory re-enters $\Omega$, at which moment we should have 
$|\bar{w}_{in}|<|\bar{w}_{out}|$. See Figure~\ref{fig:adp-illustration} for an
illustration. The proof is similar if the system starts outside $\Omega$.

Thus, ${\bf x}_1 = \left[ \begin{array}{cc} v_1 & w_1 \end{array}
\right]^T$ is always bounded, which leads to asymptotic convergence of
$\tilde{\bf x}$ to $0$ according to Theorem~\ref{th:adaptation-two}.
Moreover, since the two FN neurons synchronize along a limit cycle,
the convergence of ${\bf W}({\bf x}_2,t)\tilde{\bf a}$ to zero implies
that of $\tilde{\bf a}$.

%%%%%%%%%%%%%%%%%%%%%%%%%%%%%%%%%%%%%%%%%%%%%%%%%%%%%%%%%%%%%%%%%%%%%%%%%%%%%%%%
%
\subsection{Proof of Lemma ${\bf 1}$} 
\label{ap:negative-semi-definite-proof}
For notation simplicity, we first choose $m=1$. Since
$$
{\bf L}_{\bf \mathcal{K} \Lambda} - {\bf L}_{\bf \mathcal{K}}^2\ =\
 {\bf D}\ (\ ({\bf \mathcal{K} \Lambda})_s - 
{\bf \mathcal{K} D}^T {\bf D \mathcal{K}}\ )\ {\bf D}^T
$$
we know that $0$ is always one of its eigenvalues, with the corresponding eigenvector
$\ {\bf v} = [1, 1, \ldots, 1]^T\ $. Assume that the eigenvalues 
$\ \lambda_i({\bf L}_{\bf \mathcal{K} \Lambda})$, 
$\lambda_i({\bf L}_{\bf \mathcal{K}}^2)$,
$\lambda_i({\bf L}_{\bf \mathcal{K} \Lambda} - {\bf L}_{\bf \mathcal{K}}^2)\ $
are all arranged in increasing order for $i=1,2,\ldots,n$. 
According to Weyl's Theorem~\cite{horn},
for two Hermitian matrix ${\bf A}$ and ${\bf B}$, 
$$
\lambda_k({\bf A})+\lambda_1({\bf B})\ \le\ \lambda_k({\bf A}+{\bf B})\ \le\
\lambda_k({\bf A})+\lambda_n({\bf B})
$$
for each $k=1,2,\ldots,n$. Thus, we have
$$
\lambda_{n-k+1}({\bf L}_{\bf \mathcal{K} \Lambda} - {\bf L}_{\bf \mathcal{K}}^2)\ \le\
\lambda_n({\bf L}_{\bf \mathcal{K} \Lambda})-\lambda_k({\bf L}_{\bf \mathcal{K}}^2)
$$
which implies that, $\forall k>1$,
$\lambda_{n-k+1}({\bf L}_{\bf \mathcal{K} \Lambda} - {\bf L}_{\bf \mathcal{K}}^2)<0$
if
\begin{equation} \label{eq:condition-negative-semi}
\lambda_n({\bf L}_{\bf \mathcal{K} \Lambda})\ <\ 
               \lambda_2({\bf L}_{\bf \mathcal{K}}^2)
\end{equation}
Therefore, 
$
\lambda_n({\bf L}_{\bf \mathcal{K} \Lambda} - {\bf L}_{\bf \mathcal{K}}^2)=0
$, that is, ${\bf L}_{\bf \mathcal{K} \Lambda} - {\bf L}_{\bf \mathcal{K}}^2$
is negative semi-definite.

In fact,
$
\lambda_2({\bf L}_{\bf \mathcal{K}}^2) = \lambda_2^2({\bf L}_{\bf \mathcal{K}})
$.
Assume
$$
\max_k \lambda_{max}({\bf K}_k [{\bf \Lambda}]_k)_s\ =\ \bar{\lambda}
$$
If $\bar{\lambda} \le 0$, we have 
$\lambda_n({\bf L}_{\bf \mathcal{K} \Lambda}) \le 0$ and both the 
conditions~(\ref{eq:condition-negative-semi}) 
and~(\ref{eq:negative-semi-definite}) are always true. If
$\bar{\lambda}>0$, then
$$
\lambda_n({\bf L}_{\bf \mathcal{K} \Lambda})\ \le\  \bar{\lambda}\ \lambda_n({\bf L})
$$
and the result in Lemma~\ref{lm:negative-semi-definite} is concluded. 

In case $m>1$, 
we can follow the same proof except that zero eigenvalue here has $m$ multiplicity,
and the corresponding eigenvectors $\ \{{\bf v}_1, {\bf v}_2, \ldots, {\bf v}_m\}\ $ 
are linear combinations of the orthogonal set 
$\ [{\bf I}, {\bf I}, \ldots, {\bf I}]^T\ $ where 
$\ {\bf I} \in \mathbb{R}^{m \times m}$ is identity matrix.

%%%%%%%%%%%%%%%%%%%%%%%%%%%%%%%%%%%%%%%%%%%%%%%%%%%%%%%%%%%%%%%%%%%%%%%%%%%%%%%%
%
\subsection{Proof of Lemma ${\bf 2}$} 
\label{ap:eigenvectors-proof}
For a real symmetric matrix, the state space has an orthogonal basis consisting of all
eigenvectors. Without loss generality, we assume such an orthogonal eigenvector set
of $\ {\bf L}_{\bf \mathcal{K} \Lambda} - {\bf L}_{\bf \mathcal{K}}^2\ $ as
$\ \{{\bf v}_1, {\bf v}_2, \ldots, {\bf v}_{mn}\}$, where
$\ [{\bf v}_1, {\bf v}_2, \ldots, {\bf v}_m] = [{\bf I}, {\bf I}, \ldots, {\bf I}]^T\ $ 
are zero eigenvectors. Therefore for any vector ${\bf x}$, we have
${\bf x} = \sum_{i=1}^{mn} k_i{\bf v}_i$ and
\begin{eqnarray*}
& & {\bf x}^T\ (\ {\bf L}_{\bf \mathcal{K} \Lambda}-{\bf L}_{\bf \mathcal{K}}^2\ )\ {\bf x} \\
&=& \sum_{i=1}^{mn} k_i{\bf v}_i^T\ (\ 
    {\bf L}_{\bf \mathcal{K} \Lambda}-{\bf L}_{\bf \mathcal{K}}^2\ )\ 
    \sum_{i=1}^{mn} k_i{\bf v}_i \\
&=& \sum_{i=m+1}^{mn} k_i{\bf v}_i^T\ (\ 
    {\bf L}_{\bf \mathcal{K} \Lambda}-{\bf L}_{\bf \mathcal{K}}^2\ )\ 
    \sum_{i=m+1}^{mn} k_i{\bf v}_i \\
&=& \sum_{i=m+1}^{mn} \lambda_i k_i^2 {\bf v}_i^T {\bf v}_i
\end{eqnarray*}
where $\lambda_i$ is the $i^{th}$ eigenvalue corresponding to the eigenvector ${\bf v}_i$,
and $\ \lambda_i<0$ $\forall i>m$ for a coupled network if the 
condition~(\ref{eq:negative-semi-definite}) is true. Thus,
${\bf x}^T\ (\ {\bf L}_{\bf \mathcal{K} \Lambda}-{\bf L}_{\bf \mathcal{K}}^2\ )\ {\bf x}
=0\ $ if and only if $\ {\bf x} = \sum_{i=1}^{m} k_i{\bf v}_i\ $, that is,
$$
{\bf x}\ =\ [{\bf I}, {\bf I}, \ldots, {\bf I}]^T [k_1, \ldots, k_m]^T
       \ =\ [{\bf x}_0, {\bf x}_0, \ldots, {\bf x}_0]^T
$$
where $\ {\bf x}_0 = [k_1, \ldots, k_m]^T$. 

%%%%%%%%%%%%%%%%%%%%%%%%%%%%%%%%%%%%%%%%%%%%%%%%%%%%%%%%%%%%%%%%%%%%%%%%%%%%%%%%
%
\subsection{Positive Semi-Definite Couplings} 
\label{ap:positive-semi-definite-proof}
Assume the coupling gain of the $k^{th}$ link is
$$
{\bf K}_k = \left[ \begin{array}{cc} {\bf K}_1 & 0  \\ 0 & 0  \end{array} \right]_k
$$
where ${\bf K}_{1k}$ is symmetric positive definite and has a common dimension to all links. 
We divide the uncoupled dynamics ${\bf J}$, and in turn the block diagonal entry of 
${\bf \Lambda}$ into the form
$$
[{\bf \Lambda}]_k = \frac{\partial {\bf f}}{\partial {\bf x}} (\bar{\bf x}, t)\ =\
{\bf J}_k(\bar{\bf x}, t) \ =\ \left[ \begin{array}{cc} {\bf J}_{11} & {\bf J}_{12} \\ 
                                        {\bf J}_{21} & {\bf J}_{22}  \end{array} \right]_k
$$
where $\bar{\bf x}$ is a value between the states of two neighboring nodes ${\bf x}_i$ and 
${\bf x}_j$, and each component of ${\bf J}_k$ has the same dimension as that of the 
corresponding part in ${\bf K}_k$. Re-define the function $V$ as
$$
V = \frac{1}{2}\ (\ {\bf x}^T {\bf L}_{\bf \mathcal{K}} {\bf x}\ +\ 
                    {\bf x}^T {\bf L}_{\bf Y} {\bf x}\ +\ 
                    \tilde{\bf a}^T {\bf P}^{-1} \tilde{\bf a}\ )
$$
where $\ {\bf L}_{\bf Y} = {\bf D} {\bf Y} {\bf D}^T \ $ is a weighted Laplacian based on the
same graph as ${\bf L}_{\bf \mathcal{K}}$ but different weights
$$
{\bf Y}_k = \left[ \begin{array}{cc} 0 & 0  \\ 0 & {\bf K}_2  \end{array} \right]_k
$$
Using a modified adaptive law
$$
\dot{\hat{\bf a}}\ =\ {\bf P} {\bf W}^T({\bf x}_{\varsigma},t) 
 \sum_{j \in {\mathcal N}_{\varsigma}} ({\bf K}+{\bf Y})_{j\varsigma}\ 
 ({\bf x}_j - {\bf x}_{\varsigma})  
$$
we can show that
\begin{eqnarray*}
\dot{V} &=& {\bf x}^T\ (\ {\bf L}_{\bf (\mathcal{K}+Y) \Lambda} - 
            {\bf L}_{\bf \mathcal{K}}^2\ )\ {\bf x} \\
        &=& {\bf x}^T\ {\bf L}_{\bf (\mathcal{K}+Y) (\Lambda-\bar{\Lambda})} \ {\bf x} \ +\
            {\bf x}^T\ (\ {\bf L}_{\bf \mathcal{K} \bar{\Lambda}} - 
            {\bf L}_{\bf \mathcal{K}}^2\ )\ {\bf x} 
\end{eqnarray*}
where we define
$
[\bar{\bf \Lambda}]_k \ =\ \left[ \begin{array}{cc} \bar{\bf J}_{11} & 0 \\ 
                                                               0 & 0  \end{array} \right]_k
$
so that
$$
[{\bf \mathcal{K} \bar{\Lambda}}]_k \ =\ 
\left[ \begin{array}{cc} {\bf K}_1 \bar{\bf J}_{11} & 0 \\ 0 & 0  \end{array} \right]_k
$$
$$
[{\bf (\mathcal{K}+Y) (\Lambda-\bar{\Lambda})}]_k \ =\
\left[ \begin{array}{cc} {\bf K}_1 ({\bf J}_{11}-\bar{\bf J}_{11}) & {\bf K}_1 {\bf J}_{12} \\ 
               {\bf K}_2 {\bf J}_{21} & {\bf K}_2 {\bf J}_{22}  \end{array} \right]_k
$$
A non-positive $\dot{V}$ can be guaranteed if

$\bullet$\ ${\bf L}_{\bf \mathcal{K} \bar{\Lambda}} - {\bf L}_{\bf \mathcal{K}}^2\ \le 0\ $, 
which can be satisfied under similar condition as~(\ref{eq:negative-semi-definite});

$\bullet$\ ${\bf (\mathcal{K}+Y) (\Lambda-\bar{\Lambda})}\ < 0$, which is true if 
$\forall\ k$, the symmetric parts of $\ {\bf K}_1 ({\bf J}_{11}-\bar{\bf J}_{11})\ $ and 
$\ {\bf K}_2 {\bf J}_{22}\ $ are both negative definite, and 
$\ \sigma_{max}({\bf K}_1 {\bf J}_{12} + {\bf J}_{21}^T {\bf K}_2^T )\ $ is bounded.
An explicit condition can be derived with feedback combination analysis~\cite{wei03-2, wei03-1}.

The rest of the convergence proof are the same as that of positive definite couplings.

%%%%%%%%%%%%%%%%%%%%%%%%%%%%%%%%%%%%%%%%%%%%%%%%%%%%%%%%%%%%%%%%%%%%%%%%%%%%%%%%
%
\subsection{Leader Combination} 
\label{ap:leader-combination-proof}
Similarly to the proof of Lemma~\ref{ap:negative-semi-definite-proof},
choose $m=1$ for notational simplicity. It can be shown that $0$ is 
always one of the eigenvalues of ${\bf L}_{\bf \mathcal{K} \Lambda} - 
 {\vec{\bf L}_{\bf \mathcal{K}}}^T\ \vec{\bf L}_{\bf \mathcal{K}}$, 
which is negative semi-definite if
$$
\lambda_{n+1}({\bf L}_{\bf \mathcal{K} \Lambda})\ <\ 
 \lambda_2( {\vec{\bf L}_{\bf \mathcal{K}}}^T\ \vec{\bf L}_{\bf \mathcal{K}} )
$$ and its only eigendirection for the zero eigenvalue is ${\bf v} =
[1, 1, \ldots, 1]^T$. Since
$$
{\vec{\bf L}_{\bf \mathcal{K}}}^T\ \vec{\bf L}_{\bf \mathcal{K}} = 
\left[ \begin{array}{cc} {\bf b}^T {\bf b} & -{\bf b}^T {\bf C} \\                         
                   -{\bf C}{\bf b}  & {\bf C}^2 \end{array} \right]
$$
we have 
$$
\lambda_2( {\vec{\bf L}_{\bf \mathcal{K}}}^T\ \vec{\bf L}_{\bf \mathcal{K}} )
\ge \lambda_1( {\bf C}^2 ) = \lambda_1^2( {\bf C} )
$$
from the interlacing eigenvalues theorem for bordered matrices~\cite{horn}.
Thus a sufficient condition for synchronization is 
$$
 \lambda_1^2( {\bf C} ) > \lambda_{n+1}({\bf L}_{\bf \mathcal{K} \Lambda})
$$
This condition is equivalent to the three requirements we listed in the
first remark in Section~\ref{sec:basic-results}. Note that the connectedness
condition refers to the whole network, while the subnetwork
containing only the followers may not be connected.

If all the coupling strengths are identical with gain $\kappa$,
the synchronization condition is
$$
\kappa\ >\ \frac{\lambda_{n+1}({\bf L})\ \max_i {\bf J}_{is} }
                {\lambda_1^2(\bar{\bf L} +\mathrm{diag}\{ \gamma_i \} )}
$$
where $\bar{\bf L}$ is the Laplacian matrix for the subnetwork
containing only followers,
and ${\bf L}$ for the whole undirected leader-followers network.

The proof is similar for $m>1$.

%%%%%%%%%%%%%%%%%%%%%%%%%%%%%%%%%%%%%%%%%%%%%%%%%%%%%%%%%%%%%%%%%%%%%%%%%%%%%%%%
\section*{ACKNOWLEDGMENTS}
This work was supported in part by a grant from the National Institutes of Health.

%%%%%%%%%%%%%%%%%%%%%%%%%%%%%%%%%%%%%%%%%%%%%%%%%%%%%%%%%%%%%%%%%%%%%%%%%%%%%%%%

\end{document}